\documentclass[twocolumn]{autart}
\usepackage{natbib}
\usepackage{graphicx}

\usepackage{mathtools}
\usepackage{bm}
\usepackage{bbm}

\usepackage{graphicx}
\usepackage{amssymb}
\usepackage{xcolor}

\usepackage{algorithm}
\usepackage{algpseudocode}

\newcommand{\R}{\mathbb{R}} % reals
\newcommand{\BC}{\mathbb{C}} % 

\newcommand{\bI}{\mathbf{I}} % 
\newcommand{\bZero}{\mathbf{0}} % 

\newcommand{\BfONE}{\mathbf{1}} % all one matrix

\newcommand{\CG}{\mathcal{G}} % 
\newcommand{\CV}{\mathcal{V}} %

\newcommand{\Ker}{{\rm Ker}}

\newcommand{\CB}{\mathcal{B}}
\newcommand{\CK}{\mathcal{K}}
\newcommand{\bA}{\bar{A}}
\newcommand{\bB}{\bar{B}}
\newcommand{\Lam}{\Lambda}

\newcommand{\Span}{{\rm Span}} % column space
\newcommand{\diag}{{\rm diag}} % 
 
\newcommand{\Rank}{\text{rank}} % 
\newtheorem{example}{Example}

\newcommand*{\QE}{\hfill$\Box$}	% black square
\newenvironment{pfof}[1]{\vspace{1ex}\noindent{\textbf{Proof of
		#1:}}\hspace{0.5em}} {\hfill\QE\vspace{1ex}}

\usepackage{xcolor}
\definecolor{linkblue}{RGB}{5,112,176}
\usepackage[colorlinks=true,linkcolor=linkblue,citecolor=linkblue,urlcolor=linkblue]{hyperref}

\makeatletter
\def\ps@copyright{\let\@mkboth\@gobbletwo
  \def\@oddhead{}\let\@evenhead\@oddhead
  \def\@oddfoot{}\let\@evenfoot\@oddfoot}
\makeatother

\begin{document}

\begin{frontmatter}

\title{Charge-Balanced Control of Linear Systems: \\ Controllability, Stabilizability, and Periodic Regulation}

\thanks[footnoteinfo]{This publication was part of the project Dutch Brain
Interface Initiative (DBI2) with project number 024.005.022 of the research
programme Gravitation, financed by the Dutch Ministry of Education, Culture and
Science (OCW) via the Dutch Research Council (NWO). (Contact: yuzhen.qin@donders.ru.nl)}

\author[First,Second]{Jie Ruan},%\ead{jie.ruan@stu.ahu.edu.cn;jie.ruan@donders.ru.nl},
\author[Third]{Zonglin Liu},%\ead{z.liu@uni-kassel.de}, 
\author[Third]{Olaf Stursberg},%\ead{stursberg@uni-kassel.de},
\author[Second]{Yuan Fan},%\ead{yuanf@ahu.edu.cn},
\author[First]{Marcel van Gerven},%\ead{m.vangerven@donders.ru.nl},
\author[First]{Yuzhen Qin}%\ead{yuzhen.qin@donders.ru.nl}

\address[First]{Department of Machine Learning and Neural Computing, Donders
Institute for Brain, Cognition and Behaviour, Radboud University, Nijmegen, 6525XZ, the Netherlands}
\address[Second]{Key Laboratory of Intelligent Computing and Signal Processing of Ministry of Education, Department of Electrical Engineering and Automation, Anhui University, Hefei, 230601, China}
\address[Third]{Control and System Theory Group, Department of Electrical
Engineering and Computer Science, University of Kassel, 34121 Kassel, Germany}


\begin{abstract}
Electrical neuromodulation is a therapeutic approach that regulates neural activity by delivering electrical currents to brain tissue to treat brain disorders, such as epilepsy and Parkinson’s disease. A critical safety requirement is charge balance, which ensures zero net injected charge over each  stimulation cycle. Despite its clinical importance, the implications of charge-balanced stimulation for system dynamics and control remain poorly understood. This article investigates the controllability, stabilizability, and regulation of discrete-time linear systems under charge-balanced input constraints, motivated by applications in electrical neuromodulation. We introduce novel notions of charge-balanced controllability and stabilizability and establish necessary and sufficient conditions for both properties. We further demonstrate that nonzero steady states cannot be maintained under charge-balanced inputs, motivating the introduction of charge-balanced periodic orbits as alternative control targets. We characterize the conditions under which convergence to such orbits is achievable and develop optimal feedback controllers to stabilize the origin and track optimal charge-balanced periodic orbits. Finally, we validate the proposed control framework using a Hodgkin–Huxley neuron model, demonstrating successful suppression of tonic spiking and highlighting the potential of our approach for controlling nonlinear dynamical systems. 
\end{abstract}

\end{frontmatter}

%===============================================================================
\section{Introduction}\label{sec:intro}
Electrical neuromodulation is an effective therapeutic approach for several brain disorders, which modulates neural activity by delivering electrical pulses. For instance, deep brain stimulation can improve motor symptoms in patients with Parkinson's disease~\citep{r41} and reduce seizure frequency in drug-resistant epilepsy~\citep{r42}. However, therapeutic benefits vary across patients and often remain limited, while possibly causing side effects, e.g., speech impairment~\citep{r46} and memory problems~\citep{r45}. A main reason for these limitations is incomplete mechanistic understanding of how electrical stimulation reshapes neural dynamics to produce therapeutic effects~\citep{r21,r24}. This gap limits systematic stimulation design, and the selection of stimulation parameters still relies largely on empirical adjustment through trial and error~\citep{r44}.  %and alleviate symptoms in patients obsessive-compulsive disorder~\citep{r43}.

In the design of brain stimulation, charge balance is a fundamental safety requirement: the net charge injected to the brain over each stimulation cycle must be zero~\citep{r22}. This requirement helps mitigate the risks of irreversible electrochemical reactions, electrode degradation, and tissue damage. From a control perspective, charge balance constrains inputs across time. Yet, a systematic understanding of how this constraint shapes the system dynamics is critically missing. 

This paper therefore asks a fundamental question: \emph{How do charge-balanced control inputs affect system dynamics?} We address this question for discrete-time linear systems by investigating how charge balance affects controllability and stabilizability. The resulting analysis provides a theoretical basis for understanding the capabilities and limitations of charge-balanced control and for designing feedback controllers that meet this constraint.

\textbf{Related Work} The charge-balanced inputs can be viewed as coupled input constraints. Control of linear systems with input constraints has been extensively studied. Existing studies have considered input amplitude
bounds constraints~\citep{r4,r8}, input norm-bounded constraints~\citep{r3,r5}, and polytopic bound input constraints~\citep{r6,r7,r38}. Significant efforts have been directed toward understanding how input constraints influence controllability~\citep{r1,r2} and stabilizability of linear systems with input constraints~\citep{r4,r7,r33,r36}. Various control strategies have also been developed for linear systems with input constraints. Model predictive control has been widely used to explicitly incorporate state and input constraints~\citep{r10,r30,r31,r37}. Adaptive control methods have been developed to enforce prescribed state and input constraints for uncertain linear systems~\citep{r12,r32,r34}, while data-driven approaches have been proposed for the stabilization of linear systems subject to input constraints~\citep{r11,r35,r39}. In contrast to existing input constraints, charge balance couples control inputs across multiple sampling instants by imposing a zero-net-input requirement over each prescribed stimulation cycle. Consequently, conventional controllers designed for charge-balanced input constraints cannot be directly applied. Moreover, how such constraints affect controllability and stabilizability remains unexplored.

\textbf{Contribution.} To the best of our knowledge, this paper provides the first unified, systematic study of controllability and stabilizability for linear systems under charge-balanced inputs. The contributions are fivefold: \\
(i) We formalize these properties by defining charge-balanced controllability and stabilizability, based on the concept of a charge-balanced cycle.. \\
(ii) We derive necessary and sufficient conditions for both properties, revealing how the system's eigenvalues and the coupling between the inputs and the system's eigenmodes jointly determine whether state steering and stabilization are possible.
\\(iii) We develop two optimal feedback controllers for stabilizing the origin: a step-wise controller that updates the input at each time step, and a cycle-wise controller that computes the entire input block from a state measurement at the start of each cycle. \\(iv) We establish a fundamental limitation: a nonzero constant target cannot be maintained under charge-balanced inputs. We therefore characterize admissible charge-balanced periodic orbits and extend both feedback designs to stabilize these orbits. \\(v) Finally, simulation study of the nonlinear Hodgkin-Huxley neuron model demonstrate that a controller designed from a discrete-time linearization can successfully suppress tonic spiking, illustrating the framework’s applicability beyond linear systems.

This article substantially extends our preliminary conference paper~\citep{r40}, which addressed reachability and provided sufficient conditions for charge-balanced controllability. We establish necessary and sufficient conditions for both charge-balanced controllability and stabilizability. We further identify a fundamental limitation on maintaining nonzero constant targets and characterize admissible charge-balanced periodic orbits as alternative regulation targets. Building on these results, we develop optimal step-wise and cycle-wise feedback controllers for stabilizing the origin and tracking periodic orbits.

The remainder of this article is organized as follows. Section~\ref{sec:prob} presents the probelm formulation. Sections~\ref{sec:controllability} and~\ref{sec:cbs} construct the necessary and sufficient conditions for charge-balanced controllability and stabilizability, respectively. In Section~\ref{sec:origin}, optimal feedback controllers are designed to stabilize the origin. In Section~\ref{sec:nonzero}, optimal feedback controllers are designed to track optimal charge-balanced periodic orbits. In Section~\ref{sec:hh}, we apply the proposed controller to a nonlinear Hodgkin-Huxley neuron model to demonstrate its extended applicability. Finally, Section~\ref{sec:con} concludes this article.

%===============================================================================
\textbf{Notation:} $\R$ and $\BC$ denote the real and complex numbers. For a matrix $M$, we write $\sigma(M)$ for its spectrum, $\Span(M)$ for its column space, and $\Ker(M)$ for its null space. For a symmetric matrix $W$, $W\succ0$ indicates that matrix $W$ is positive definite. $\bI_n$ is the $n\times n$ identity, 
$\bZero$ is a zero vector or matrix of the size implied by the context, and $\BfONE_h$ is the all-ones vector of length $h$. Eigenvectors are taken in $\BC^n$; by a \emph{left eigenvector} of $A$ associated with $\lambda\in\sigma(A)$ we mean a nonzero $\phi\in\BC^n$ with $\phi^\top A=\lambda\phi^\top$, equivalently
$\phi\in\Ker(A^\top-\lambda \bI_n)$.  We write $\CK_r(A,B):=\Span\big[\,B\ \ AB\ \ \cdots\ \ A^{r-1}B\,\big]$ for the $r$-step reachable subspace of $(A,B)$; the reachable subspace of the pair is $\CK_n(A,B)$, and the \emph{controllability index} is $\nu:=\min\{r\ge1:\CK_r(A,B)=\R^n\}$ when $(A,B)$ is controllable.  
\section{Problem formulation}\label{sec:prob}
Consider a discrete-time linear system described by 
\begin{align}\label{eq:sys}
    &x_{k+1} = A x_{k} +Bu_{k},&k=0,1, 2, \dots,
\end{align}
where $x_k\in \R^n, u_k\in\R^m$. Motivated by the notion of a stimulation cycle, we make the following assumption to ensure that the control inputs satisfy the charge-balanced condition commonly required in electrical brain neurostimulation paradigms.

\begin{assum}[Charge balance]\label{as:cb}
There is an integer $h\ge2$ (the \emph{cycle length}) such that the input
satisfies
\begin{equation}\label{eq:cb}
  \sum\nolimits_{i=0}^{h-1}u_{ph+i}=\bZero,\qquad p=0,1,2,\dots.
\end{equation}
\end{assum}

It states that within each cycle the positive and negative inputs injected to the system cancel.  The cycle length $h$ is a design parameter. Since charge balance is usually achieved over a short time window, small $h$ is preferred.  Note that $h=1$ would force $u\equiv\bZero$,  $h\ge2$ is thus assumed.
For the $p$-th cycle, collect the inputs into
\begin{equation*}
	U_p:=\big[u_{ph}^\top\ \ u_{ph+1}^\top\ \ \cdots\ \ u_{(p+1)h-1}^\top\big]^\top \in\R^{mh}
\end{equation*}
and define $R:=\BfONE_h^\top\otimes \bI_m\in\R^{m\times mh}$. Then, Assumption~\ref{as:cb} can be rewritten as
\begin{align}\label{input_constrain}
    &RU_p = \bZero, &\forall p=0, 1, 2, \dots.
\end{align}

In this article, our goals are twofold. First, we aim to characterize the conditions under which the system~\eqref{eq:sys} is controllable with charge-balanced inputs. Then, we want to design charge-balanced controllers to stabilize and regulate this system. 

\section{Controllability with charge-balanced inputs}\label{sec:controllability}
In this section, we study the controllability problem, which is defined as follows.

\begin{defn}[Charge-balanced controllability]\label{def:cbc}
For a fixed $h\ge2$, system~\eqref{eq:sys} is said to be
\emph{$h$-charge-balanced controllable} ($h$-CBC) if, for every
$x_0,x_f\in\R^n$, there exist a finite integer $b\ge1$ and a sequence
of input blocks $U_0,U_1,\ldots,U_{b-1}$ satisfying
\eqref{input_constrain} such that $x_{bh}=x_f$.
\end{defn} 

Note that terminal states are set at cycle boundaries so
that the input sequence actually applied is charge balanced
as a whole. 

To characterize $h$-CBC, we lift system~\eqref{eq:sys} at the cycle boundaries.
Define the matrix $Q\in\R^{mh\times m(h-1)}$ with columns forming the basis of $\Ker(R)$, i.e., $\Span(Q)=\Ker(R)$. Without loss of generality, one can select $Q$ with orthonormal columns. Then, it holds that $Q^\top Q=I_{m(h-1)}$. Consequently, it follows that control inputs within each charge-balanced block $p$ satisfy
\begin{align}\label{def_Q}
     U_p=Qw_p,
\end{align}
where $w_p\in\R^{m(h-1)}$ is a free input. Now, define
\begin{equation}\label{eq:S}
  S\coloneqq\big[\,A^{h-1}B\ \ A^{h-2}B\ \ \cdots\ \ AB\ \ B\,\big]\in\R^{n\times mh}.
\end{equation}
Then, one can derive a \textit{lifted system} based on charge-balanced blocks: 
\begin{align}\label{eq:lift}
    x_{(p+1)h} =\bar A x_{ph}+\bar B w_p,
\end{align}
where $\bar A \coloneqq  A^h\in\R^{n\times n}$ and $\bar B\coloneqq SQ\in\R^{n\times m(h-1)}$. 

The following Lemma reformulates the $h$-CBC problem of system~\eqref{eq:sys} as a controllability problem for the lifted system~\eqref{eq:lift}, whose proof is given in the Appendix.

\begin{lem}\label{lem:equiv} 
System~\eqref{eq:sys} is $h$-CBC if and only if the pair
$(\bA,\bB)=(A^h,SQ)$ is controllable, independent of the particular choice of the basis matrix $Q$.
\end{lem}

Therefore, it remains to characterize the controllability of the lifted pair $(\bA,\bB)$. We first identify the effective input space induced by charge balance. 

\begin{lem}[Effective input space]\label{lem:image}
For every $h\ge2$, it holds that
\begin{equation}\label{eq:image}
\begin{split}
  \Span(\bB)&={\textstyle\sum_{j=1}^{h-1}}\Span\big((A^{j}-\bI_n)B\big)\\
            &=(A-\bI_n)\,\CK_{h-1}(A,B).
\end{split}
\end{equation}
\end{lem}

Lemma~\ref{lem:image}  shows how charge balance acts on the system, which distinguishes from the situation without the charge-balanced constraint, where the right-hand side of~\eqref{eq:image} is $\CK_h(A,B)$. This lemma will play a very important role in later analysis.

Before presenting the main result of this section, we provide some further definitions.

\begin{figure}[t]
	\centering
	\begin{minipage}[c]{0.5\linewidth}
		\centering
		\includegraphics[width=0.8\linewidth]{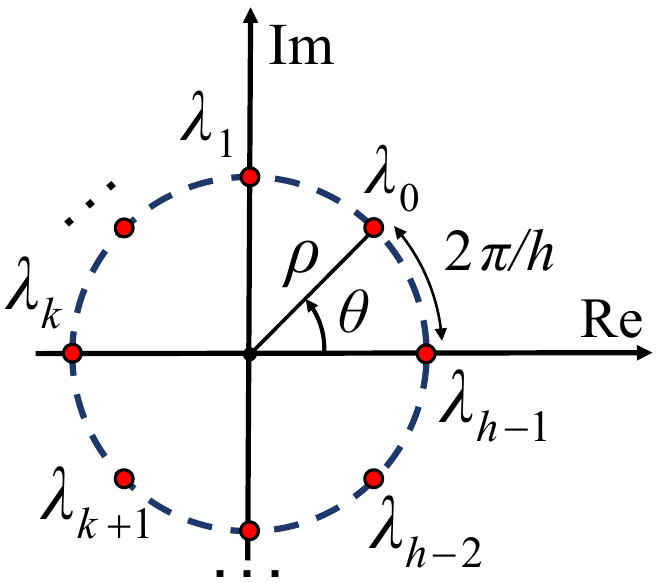}
	\end{minipage}\hfill
	\begin{minipage}[c]{0.46\linewidth}
		\centering\small
		\textbf{Example}
		\par \vspace{4pt}
		$\begin{gathered}
		A=\begin{bmatrix}0&2&0\\0&0&2\\2&0&0\end{bmatrix}\\[5pt]
		\begin{aligned}
		\lambda_1=2,
		\lambda_{2,3}=-1 \pm \eta\sqrt{3},
		\end{aligned}
		\end{gathered}$
	\end{minipage}
	\caption{Left: illustration of Definition~\ref{def:cycle}. Right: a matrix example, whose eigenvalues form a complete spectral $3$-cycle, as  they satisfy $\lambda_i^3=8$ (here $\eta$ is the imaginary unit).}
	\label{h_cycle}
\end{figure}

\begin{defn}[Complete spectral $h$-cycle]\label{def:cycle}
A discrete set $\Lam$ is a \emph{complete spectral $h$-cycle} if there is
$\mu\in\BC\setminus\{0\}$ such that $\Lam=\{z\in\BC:z^h=\mu\}$, equivalently
$\Lam=\{\rho\,e^{\eta(\theta+2\pi k/h)}:k=0,\dots,h-1\}$ for some $\rho>0$ and
$\theta\in\R$.  Thus $\Lam$ consists of exactly $h$ distinct points equally spaced on a circle centered at the origin in the complex plane (see Fig.~\ref{h_cycle}).
\end{defn}

\begin{defn}[Modal input space]\label{def:modal}
For $\lambda\in\sigma(A)$ the \emph{modal input space} of $\lambda$ is
\begin{equation*} 
  \CB_\lambda:=\big\{\phi^\top B:\ \phi\in\Ker(A^\top-\lambda\bI_n)\big\}
  \subseteq\BC^{1\times m}.
\end{equation*}
\end{defn}

The space $\CB_\lambda$ collects the row vectors that describe how the inputs act on the left modal coordinates associated with $\lambda$.

\begin{thm}[Controllability]\label{thm:CBC}
Let $h\ge2$ and assume $h\ge\nu_0+1$, where $\nu_0$ is the smallest nonnegative integer\footnote{Such $\nu_0$ is also called the index of the eigenvalue $0$ of $A$. If $A$ is invertible, $\nu_0=0$.} such that $\ker(A^k)=\ker(A^{k+1})$.  Then system~\eqref{eq:sys} is $h$-CBC if and only if
\begin{enumerate}
  \item[\textup{(i)}] the pair $(A,B)$ is controllable;
  \item[\textup{(ii)}] $1\notin\sigma(A)$; and
  \item[\textup{(iii)}] for any subset $\Lam\subseteq\sigma(A)$ that is a complete spectral $h$-cycle, it satisfies  $\cap_{\lambda\in\Lam}\CB_\lambda=\{\bZero\}$.
\end{enumerate}

\end{thm}

Note that if no such subset $\Lambda$ exists, condition~\textup{(iii)} is said to hold vacuously. Condition (i) rules out the situations where the original system is uncontrollable.  Condition (ii) excludes the eigenvalue $\lambda=1$; otherwise, we have $\phi^\top A=\phi^\top$, then 
\begin{equation*}
\phi^\top x_{(p+1)h}
=\phi^\top x_{ph}+\phi^\top B\sum_{i=0}^{h-1}u_{ph+i}
=\phi^\top x_{ph},
\end{equation*}
so this eigenmode is conserved at every cycle boundary under any charge-balanced input and  cannot be steered. Condition~(iii) prevents a loss of controllability caused by the lifting $A\mapsto A^h$. The eigenvalues in a complete spectral $h$-cycle become the same eigenvalue of $A^h$. The condition requires that these input spaces do not all share a common nonzero input direction. Otherwise, charge balance can make a linear combination of them uncontrollable.

\subsection{Intermediate results}

Before proving Theorem~\ref{thm:CBC}, we provide some instrumental results, whose proofs can be found in Appendix~\ref{proof_sec_3}.

\begin{lem}[Necessary conditions]\label{lem:necessary}
If the system~\eqref{eq:sys} is $h$-CBC, then it holds that: (i) $(A,B)$ is controllable, and (ii) $1\notin\sigma(A)$. 
\end{lem}

\begin{lem}[Eigenspaces after lifting]\label{lem:lift}
Let $\mu\in\BC\setminus\{0\}$ and set $\Lam_\mu:=\{\lambda\in\sigma(A):\lambda^h=\mu\}$.  Then 
\begin{equation}\label{eq:lifteig}
  \Ker\big((A^\top)^h-\mu \bI_n\big)
  =\oplus_{\lambda\in\Lam_\mu}\Ker(A^\top-\lambda \bI_n).
\end{equation}
Here $\oplus$ is the direct sum, i.e., every $\phi$ in the left-hand side has a \emph{unique} decomposition $\phi=\sum_{\lambda\in\Lam_\mu}\phi_\lambda$ with
$\phi_\lambda\in\Ker(A^\top-\lambda \bI_n)$. 
\end{lem}
To prove Theorem~\ref{thm:CBC}, by Lemma~\ref{lem:equiv}, it suffices to examine the PBH controllability test for the lifted pair $(\bA,\bB)$. Therefore, we investigate whether there exists a left eigenvector $\phi\neq\bZero$ and $\mu\in\BC$ such that $\phi^\top A^h=\mu\phi^\top$ and $\phi^\top\bar B=\bZero^\top$. For $\mu\neq0$, Lemma~\ref{lem:lift} implies that every such left eigenvector $\phi$ decomposes uniquely into modal components associated with the roots of $\lambda^h=\mu$. Consequently, the PBH test on the lifted system naturally splits into three distinct cases according to the nature of the lifted eigenvalue $\mu$: (i) $\mu=0$, (ii) $\mu\neq0$ with $|\Lambda_\mu|\le h-1$, and (iii) $\mu\neq0$ with $|\Lambda_\mu|= h$, forming a \textit{complete spectral $h$-cycle}. The following three lemmas address these scenarios respectively.
\begin{lem}[Zero eigenvalue]\label{lem:zero}
Assume $(A,B)$ is controllable and $h\ge\nu_0+1$.  Then there is no $\phi\in\BC^n\setminus\{\bZero\}$ such that $\phi^\top A^h=\bZero^\top$ and
$\phi^\top\bB=\bZero^\top$.
\end{lem}

\begin{lem}[Incomplete spectral $h$-cycles]\label{lem:partialcycle}
Assume that $(A,B)$ is controllable and $1\notin\sigma(A)$. Let
$\mu\in\BC\setminus\{0\}$ and
$\Lambda_\mu:=\{\lambda\in\sigma(A):\lambda^h=\mu\}$.
If $|\Lambda_\mu|\le h-1$, then there is no nonzero
$\phi\in\Ker((A^\top)^h-\mu \bI_n)$ such that
$\phi^\top\bar B=\bZero^\top$.
\end{lem}

\begin{lem}[Complete spectral $h$-cycles]\label{lem:completecycle}
Assume that $(A,B)$ is controllable and $1\notin\sigma(A)$. Let
$\mu\in\BC\setminus\{0\}$ and
$\Lambda_\mu:=\{\lambda\in\sigma(A):\lambda^h=\mu\}$.
If $\Lambda_\mu$ is a complete spectral $h$-cycle, then there exists a nonzero
$\phi\in\Ker((A^\top)^h-\mu \bI_n)$ satisfying
$\phi^\top\bar B=\bZero^\top$ if and only if $\bigcap_{\lambda\in\Lambda_\mu}\CB_\lambda\neq\{\bZero\}$.
\end{lem}

\subsection{Proof of Theorem~\ref{thm:CBC}}

\emph{Sufficiency.}
Assume \textup{(i)}-\textup{(iii)}. By Lemma~\ref{lem:equiv}, it suffices to show that
$(\bar A,\bar B)$ is controllable. Suppose, to the contrary, that $(\bar A,\bar B)$ is uncontrollable. By the PBH test, there exist $\mu\in\BC$ and $\phi\neq\bZero$ such that
$\phi^\top A^h=\mu\phi^\top$ and $\phi^\top\bar B=\bZero^\top$. If $\mu=0$, this contradicts
Lemma~\ref{lem:zero}. Hence $\mu\neq0$. Let $\Lambda_\mu :=\{\lambda\in\sigma(A) :\lambda^h=\mu\}$ and $s:=|\Lambda_\mu|$. Since $\mu\in\sigma(A^h)$, $1\le s\le h$. If $s\le h-1$, Lemma~\ref{lem:partialcycle} gives a contradiction. Thus, $s=h$, so $\Lambda_\mu$ is a complete spectral $h$-cycle. By Lemma~\ref{lem:completecycle}, $\bigcap_{\lambda\in\Lambda_\mu}\CB_\lambda\neq\{\bZero\}$, which contradicts \textup{(iii)}. Therefore, $(\bar A,\bar B)$ is controllable, and system~\eqref{eq:sys} is $h$-CBC.

\emph{Necessity.}
Suppose that system~\eqref{eq:sys} is $h$-CBC. By Lemma~\ref{lem:necessary}, conditions \textup{(i)} and \textup{(ii)} hold. It remains to prove \textup{(iii)}.
Suppose, to the contrary, that \textup{(iii)} fails. Then there exists a complete spectral $h$-cycle $\Lambda_\mu\subseteq\sigma(A)$ such that $ \bigcap_{\lambda\in\Lambda_\mu}\CB_\lambda\neq\{\bZero\}$. Since \textup{(i)} and \textup{(ii)} hold, Lemma~\ref{lem:completecycle} implies that there exists $\phi\neq\bZero$ satisfying $\phi^\top A^h=\mu\phi^\top$ and $\phi^\top\bar B=\bZero^\top$. Hence $(\bar A,\bar B)$ fails the PBH test and is uncontrollable, contradicting
Lemma~\ref{lem:equiv}. Therefore, \textup{(iii)} holds.
\hfill$\Box$

\section{Stabilizability with charge-balanced inputs}\label{sec:cbs}
In this section, we characterize the stabilizability of the system~\eqref{eq:sys} under charge-balanced control.

\begin{defn}[Charge-balanced stabilizability]\label{def:cbs}
For a fixed $h\ge2$, system~\eqref{eq:sys} is said to be \emph{$h$-charge-balanced stabilizable} ($h$-CBS) if, for every $x_0\in\R^n$, there exists an input sequence satisfying \eqref{input_constrain} such that $\lim_{k\to\infty}x_k=\bZero$.
\end{defn}

The following lemma relates $h$-CBS to stabilizability of the lifted system~\eqref{eq:lift}. Its proof is given in the Appendix.

\begin{lem}\label{lem:cbs_equiv}
Let $h\ge2$. System~\eqref{eq:sys} is $h$-CBS if and only if the pair $(\bA,\bB)=(A^h,SQ)$ is stabilizable, independent of the particular choice of the basis matrix $Q$.
\end{lem}

\begin{thm}[Stabilizability]\label{thm:CBS}
Let $h\ge2$. System~\eqref{eq:sys} is $h$-CBS if and only if
\begin{enumerate}
  \item[\textup{(i)}] the pair $(A,B)$ is stabilizable;
  \item[\textup{(ii)}] $1\notin\sigma(A)$; and
  \item[\textup{(iii)}] for any subset $\Lam\subseteq\sigma(A)$ that
  is a complete spectral $h$-cycle with $|\lambda|\ge1$ for all
  $\lambda\in\Lam$, it satisfies
  $\cap_{\lambda\in\Lam}\CB_\lambda=\{\bZero\}$.
\end{enumerate}
\end{thm}

If no subset $\Lam$ as in \textup{(iii)} exists, condition \textup{(iii)} is understood to hold vacuously. The proof is given in the Appendix~\ref{proof_sec_4}. Theorem~\ref{thm:CBS} is the stabilization counterpart of Theorem~\ref{thm:CBC}, with less restrictive conditions. Conditions \textup{(i)} and \textup{(iii)} of Theorem~\ref{thm:CBS} relax Conditions~\textup{(i)} and \textup{(iii)} of Theorem~\ref{thm:CBC} to the cases with $|\lambda|\ge1$. The assumption $h\ge\nu_0+1$ is no longer needed, because the zero-eigenvalue part of $A$ is stable. Since controllability of $(\bA,\bB)$ implies its stabilizability, Lemmas~\ref{lem:equiv} and~\ref{lem:cbs_equiv} show that $h$-CBC implies $h$-CBS, but not the converse.

\section{Stabilization at the origin ($x_f=\bZero$)}\label{sec:origin}

This section designs charge-balanced controllers that stabilize an $h$-CBS system. 

First, we define a quadratic cost function. To this end, define the accumulated charge by
\begin{equation}\label{eq:charge_state}
q_0:=\bZero,\qquad q_{k+1}:=q_k+u_k,\qquad k=0,1,2,\dots.
\end{equation}
Since $q_{(p+1)h}-q_{ph}=\sum_{i=0}^{h-1}u_{ph+i}$, charge balance~\eqref{eq:cb} holds if and only if $q_{ph}=\bZero$ for every $p$. In that case, $q_{ph+i}=\sum_{j=0}^{i-1}u_{ph+j}$ is the net charge delivered so far in the current cycle, and the last input of every cycle equals $u_{ph+h-1}=-q_{ph+h-1}$. The cost function is defined by 
\begin{align}\label{eq:lqr_cost}
J=\sum_{p=0}^{\infty}\sum_{i=0}^{h-1}
\big(&x_{ph+i}^\top W_{x,i}x_{ph+i}
+u_{ph+i}^\top W_uu_{ph+i}\notag\\
&+q_{ph+i}^\top W_qq_{ph+i}\big),
\end{align}
where $W_{x,0}\succ0$, $W_{x,i}\succeq0$ for $i=1,\ldots,h-1$, $W_u\succ0$, and $W_q\succeq0$. The matrices $W_{x,i}$ penalize state deviations from the origin at each phase, while $W_u$ penalizes control effort. The optional weight $W_q$ penalizes temporary charge accumulation within a cycle, supplementing the requirement of zero net charge at cycle boundaries. Because the last input of every cycle is determined by the previous ones, minimizing~\eqref{eq:lqr_cost} subject to~\eqref{input_constrain} is not a standard LQR problem for $(A,B)$.

Next, we design two types of controllers.

\subsection{Optimal step-wise feedback control}
Define the augmented state $\xi_k:=[\,x_k^\top\ \ q_k^\top\,]^\top\in\R^{n+m}$, which evolves as
\begin{equation}\label{eq:aug}
\xi_{k+1}=\tilde A\xi_k+\tilde Bu_k,\quad
\tilde A:=\begin{bmatrix}A&\bZero\\ \bZero&\bI_m\end{bmatrix},\quad
\tilde B:=\begin{bmatrix}B\\ \bI_m\end{bmatrix}.
\end{equation}
The inputs at phases $0,\ldots,h-2$ are free, whereas the last input is fixed at $-q_{ph+h-1}$. Hence, the last step of each cycle is $\xi_{(p+1)h}=\tilde A_{\rm e}\xi_{ph+h-1}$ with $\tilde A_{\rm e}:=\tilde A-\tilde B[\,\bZero\ \ \bI_m\,]$, and the charge-balanced system is an $h$-periodic system in $\xi$. The stage cost becomes $\xi^\top\tilde W_i\xi+u^\top W_uu$ at phases $i\le h-2$ and $\xi^\top\tilde W_{\rm e}\xi$ at phase $h-1$, where
$\tilde W_i:=\diag(W_{x,i},W_q)$ and $\tilde W_{\rm e}:=\diag(W_{x,h-1},W_q+W_u)$.

 Let $V_i(\xi)=\xi^\top\Pi_i\xi$ denote the optimal remaining cost at phase $i$, with $V_h=V_0$. At phases $i=0,\ldots,h-2$, minimizing the stage cost plus $V_{i+1}(\tilde A\xi+\tilde Bu)$ gives $u=-\tilde K_i\xi$. At phase $h-1$, the input is fixed at $u=-q$. Substituting these inputs into the Bellman equations yields the periodic Riccati equations for the unknown symmetric matrices $\Pi_0,\ldots,\Pi_{h-1}\succeq0$:
\begin{align}\label{eq:sweep}
\Pi_{h-1}&=\tilde W_{\rm e}+\tilde A_{\rm e}^\top\Pi_0\tilde A_{\rm e},\notag\\
\Pi_i&=\tilde W_i+\tilde A^\top\Pi_{i+1}\tilde A
-\tilde K_i^\top\Theta_i\tilde K_i,
\end{align}
for $i=h-2,\ldots,0$, where
\begin{equation}\label{eq:step_gain}
\Theta_i:=W_u+\tilde B^\top\Pi_{i+1}\tilde B,\qquad
\tilde K_i:=\Theta_i^{-1}\tilde B^\top\Pi_{i+1}\tilde A .
\end{equation}
Denote $[\Pi]_x$ the leading $n\times n$ block of $\Pi$.
Since $\tilde A_{\rm e}\xi$ has zero charge, only $[\Pi_0]_x$
enters the first equation. The optimal step-wise controller is
\begin{equation}\label{eq:step_controller}
u_{ph+i}=
\begin{cases}
-\tilde K_i\xi_{ph+i}, & i=0,\ldots,h-2,\\
-q_{ph+h-1}, & i=h-1.
\end{cases}
\end{equation}
The gains are the same for each cycle. The controller uses the current state and the charge $q_k$, computed from past inputs, and the final input ensures charge balance.

\begin{thm}[Step-wise feedback]\label{thm:step}
Suppose that system~\eqref{eq:sys} is $h$-CBS. Then,
\begin{enumerate}
\item[\textup{(i)}] the Riccati equations \eqref{eq:sweep} has a unique positive semidefinite solution;
\item[\textup{(ii)}] for every $x_0\in\R^n$, the step-wise controller~\eqref{eq:step_controller} achieves the minimum $x_0^\top[\Pi_0]_xx_0$ of~\eqref{eq:lqr_cost} over all charge-balanced control inputs;
\item[\textup{(iii)}] under the step-wise controller~\eqref{eq:step_controller}, the origin is globally  asymptotically stable.
\end{enumerate}
\end{thm}

\subsection{Optimal cycle-wise feedback control}\label{sec:cycle_control}
The step-wise controller requires a state measurement at every step. A simpler implementation measures the state only at the start of each cycle and fixes all $h$ inputs of the cycle at that time, i.e., 
\begin{equation}\label{eq:original_block_feedback}
u_{ph+i}=L_ix_{ph},\qquad i=0,\ldots,h-1,
\end{equation}
where $L_i\in\R^{m\times n}$. Charge balance for every $x_{ph}$ requires $\sum_{i=0}^{h-1}L_i=\bZero$, i.e., $R\mathcal L=\bZero$ for $\mathcal L:=[\,L_0^\top\ \cdots\ L_{h-1}^\top\,]^\top$. Since the columns of $Q$ form a basis of $\Ker(R)$, $\mathcal L=QK$ for a unique $K\in\R^{m(h-1)\times n}$. Partitioning $Q=[\,Q_0^\top\ \cdots\ Q_{h-1}^\top\,]^\top$ into block rows gives $L_i=Q_iK$, so the controller applies $U_p=QKx_{ph}$, i.e., $w_p=Kx_{ph}$. Define $S_0:=\bZero_{n\times mh}$ and
\begin{equation*}
S_i:=\big[\,A^{i-1}B\ \cdots\ B\ \ \bZero_{n\times m(h-i)}\,\big],
\quad i=1,\ldots,h-1.
\end{equation*}
Then, for $i=0,\ldots,h-1$,
\begin{align}\label{d10}
x_{ph+i}=A^ix_{ph}+S_iQw_p
=(A^i+S_iQK)x_{ph},
\end{align}
and the boundary states satisfy
\begin{equation}\label{eq:closed}
x_{(p+1)h}=Fx_{ph},\qquad F:=\bA+\bB K.
\end{equation}
Since there are finitely many phases, $\|x_{ph+i}\|\le c\|x_{ph}\|$ for all $i$, with $c:=\max_{i}\|A^i+S_iQK\|$. Hence the block controller makes the origin globally asymptotically stable if and only if $F$ is Schur.

To select $K$, we express the cost of one cycle in terms of $x_{ph}$ and $w_p$. Since $q_{ph}=\bZero$ for charge-balanced inputs, within the $p$th cycle $q_{ph+i}=E_iU_p$, where $E_i:=[\,\BfONE_i^\top\otimes\bI_m\ \ \bZero_{m\times m(h-i)}\,]$ for $i\ge1$ and $E_0:=\bZero$. Substituting~\eqref{d10}, $U_p=Qw_p$, and $q_{ph+i}=E_iQw_p$ into~\eqref{eq:lqr_cost} gives
\begin{equation}\label{eq:lifted_cost}
J=\sum_{p=0}^{\infty}\ell(x_{ph},w_p),\quad
\ell(x,w):=
\begin{bmatrix}x\\ w\end{bmatrix}^{\!\top}\!
\begin{bmatrix}{ \bar W_x}&{ \bar W_{xw}}\\ { \bar W_{xw}}^\top&{ \bar W_w}\end{bmatrix}
\begin{bmatrix}x\\ w\end{bmatrix},
\end{equation}
where
\begin{align*}
{ \bar W_x}&:=\sum_{i=0}^{h-1}(A^i)^\top W_{x,i}A^i,\qquad
{ \bar W_{xw}}:=\sum_{i=0}^{h-1}(A^i)^\top W_{x,i}S_iQ,\\
{ \bar W_w}&:=Q^\top\Big(\bI_h\otimes W_u
+\sum_{i=0}^{h-1}\big(S_i^\top W_{x,i}S_i+E_i^\top W_qE_i\big)\Big)Q.
\end{align*}
In general, ${ \bar W_{xw}}\neq0$, because penalizing the intra-cycle states introduces a cross term between the boundary state and the input block. If $W_{x,i}=\bZero$ for $i\ge1$, then ${ \bar W_x}=W_{x,0}$ and ${ \bar W_{xw}}=\bZero$, regardless of $W_q$. Thus, \eqref{eq:lifted_cost} defines an LQR problem for $(\bA,\bB)$ with a cross term. Let $P$ be the stabilizing solution of the discrete algebraic Riccati equation
\begin{align}\label{eq:lqr_dare}
P={}&\bA^\top P\bA+{ \bar W_x}-(\bA^\top P\bB+{ \bar W_{xw}})\notag\\
&\times({ \bar W_w}+\bB^\top P\bB)^{-1}(\bB^\top P\bA+{ \bar W_{xw}}^\top),
\end{align}
and define
\begin{equation}\label{eq:optimal_K}
K^*:=-({ \bar W_w}+\bB^\top P\bB)^{-1}(\bB^\top P\bA+{ \bar W_{xw}}^\top).
\end{equation}
The optimal cycle-wise controller controller is
\begin{equation}\label{eq:lqr_original_controller}
U_p=QK^*x_{ph}.
\end{equation}

\begin{thm}[Cycle-wise feedback]\label{thm:block}
Suppose that system~\eqref{eq:sys} is $h$-CBS. Then,
\begin{enumerate}
\item[\textup{(i)}] \eqref{eq:lqr_dare} has a unique stabilizing solution $P\succeq0$;
\item[\textup{(ii)}] for every $x_0\in\R^n$, the cycle-wise
controller~\eqref{eq:lqr_original_controller} achieves the minimum
$x_0^\top Px_0$ of~\eqref{eq:lqr_cost} over all charge-balanced control inputs;
\item[\textup{(iii)}] under the cycle-wise
controller~\eqref{eq:lqr_original_controller}, the origin is globally asymptotically stable.
\end{enumerate}
\end{thm}

The proof is given in Appendix~\ref{proof_sec_5_6}.
For each fixed $x_0$, the minimizing input sequence is unique: the charge-balance constraints define a convex set, and $W_u\succ0$ makes the cost strictly convex on the admissible input sequences with finite cost. By Theorems~\ref{thm:step} and~\ref{thm:block}, both controllers achieve this minimum and therefore generate identical input and state trajectories from the same initial state. Their optimal values coincide for every $x_0$, so $P=[\Pi_0]_x$.

The step-wise controller uses the current state and accumulated charge at phases $0,\ldots,h-2$, and cancels the remaining charge at the last phase. The cycle-wise controller computes the entire input block $U_p=QK^*x_{ph}$ at the start of each cycle and applies it without further state measurements during that cycle. Its gain follows from the standard Riccati equation~\eqref{eq:lqr_dare}. Note that step-wise controller over one cycle is $U_p=\mathcal L_{\rm sw}x_{ph}$, where $R\mathcal L_{\rm sw}=\bZero$. Hence, there exists a unique matrix $K_{\rm sw}\in\R^{m(h-1)\times n}$ such that $\mathcal L_{\rm sw}=QK_{\rm sw}$ and $U_p=QK_{\rm sw}x_{ph}$. Since the step-wise and the cycle-wise controllers minimize the same cost function~\eqref{eq:lqr_cost}, whose optimal input sequence is unique, they have the same optimal input for any given initial condition.

\begin{figure}[t]
\centering
\includegraphics[width=0.9\columnwidth]{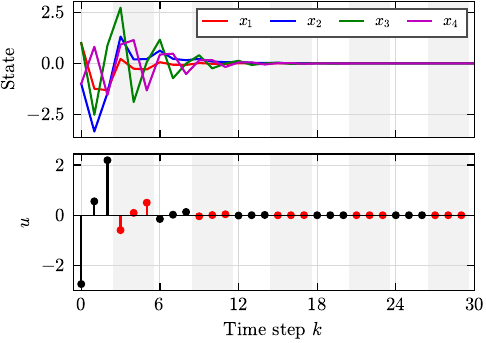}
\caption{Block control of a $3$-CBS system that is not $3$-CBC.
Shaded and unshaded bands mark alternate cycles; the inputs of each cycle sum to zero.}
\label{fig:cbs_stab}
\end{figure}

\begin{example}
    Consider system~\eqref{eq:sys} with $h=3$,
\begin{equation*}
A=\begin{bmatrix}
1.5&0&0&0\\
0&0.6&0&0\\
0&0&-0.3&-0.3\sqrt3\\
0&0&0.3\sqrt3&-0.3
\end{bmatrix}, \qquad
B=\begin{bmatrix}1\\ 1\\ 1\\0 \end{bmatrix},
\end{equation*}
The pair $(A,B)$ is controllable, and the eigenvalues of $A$ are $1.5$, $0.6$, and $0.6e^{\pm\eta2\pi/3}$, so Conditions \textup{(i)} and \textup{(ii)} hold. The only complete spectral $3$-cycle in $\sigma(A)$ is
$\{0.6,\,0.6 e^{\pm\eta2\pi/3}\}$, which lies inside the unit disk, so Condition \textup{(iii)} holds and the system is $3$-CBS by Theorem~\ref{thm:CBS}. Since $m=1$, $\CB_\lambda=\BC$ for every eigenvalue, so this cycle violates Condition \textup{(iii)} and the system is not $3$-CBC. The block controller~\eqref{eq:lqr_original_controller} is designed with $W_{x,i}=\bI_4$, $W_u=1$, and $W_q=0$.
Fig.~\ref{fig:cbs_stab} shows the trajectories starting from $x_0=[1~\ -1~\ 1~\ -1]^\top$.
\end{example}

\section{Regulation towards nonzero target ($x_f\neq\bZero$)}\label{sec:nonzero}
In this section, we consider a nonzero target $x_f\neq \bZero$. By Definition~\ref{def:cbc}, if system~\eqref{eq:sys} is $h$-CBC, $x_f$ can be reached with charge-balanced control. However, we next show that the state cannot remain at $x_f$ at all time steps.

\begin{thm}[Nonzero targets cannot be held]\label{thm:nohold}
Suppose that $1\notin\sigma(A)$, and let $\{x_k\}$ be generated by system~\eqref{eq:sys} with any input sequence satisfying~\eqref{input_constrain}. Then, for every cycle $p\ge0$,
\begin{equation}\label{eq:cycle_sum}
(\bI_n-A)\sum_{i=0}^{h-1}x_{ph+i}=x_{ph}-x_{(p+1)h}.
\end{equation}
Consequently, $x_k\to x_f$ implies $x_f=\bZero$. In particular, the state cannot remain at any $x_f\neq\bZero$.
\end{thm}

The proof can be found in Appendix~\ref{proof_sec_5_6}. Theorem~\ref{thm:nohold} applies to every charge-balanced input sequence. It follows from the factor $A-\bI_n$ of Lemma~\ref{lem:image}, i.e., the inputs of each cycle sum to zero, so they drop out of~\eqref{eq:cycle_sum}, and the sum of the states over a cycle is determined by the change of the boundary state alone. If the change is $0$, i.e., $x_{ph}-x_{(p+1)h}=\bZero$, the mean state over the cycle is zero. This state structure motivates us to investigate the charge-balanced $h$-periodic orbit with definition given below.
\begin{defn}[Charge-balanced $h$-periodic orbit]\label{def:cb_periodic_orbit}
A pair $\mathcal{O}:=\left(\{z_i\}_{i=0}^{h-1},\{u_i\}_{i=0}^{h-1}\right)$
is called a charge-balanced $h$-periodic orbit of
system~\eqref{eq:sys} if: (i) $z_{i+1}=Az_i+Bu_i,i=0,\ldots,h-1$, and (ii) $\sum_{i=0}^{h-1}u_i=\bZero$, where the state indices are interpreted modulo $h$, so that $z_h=z_0$.
\end{defn}

Since the boundary state of an orbit repeats,~\eqref{eq:cycle_sum}
shows that every orbit has zero mean, i.e.,
\begin{equation}\label{eq:zero_sum}
\sum\nolimits_{i=0}^{h-1}z_i=\bZero .
\end{equation}

Notice that each orbit is determined by its boundary state $z_0$ and its input
block $U_f=Qw_f$, $w_f\in\R^{m(h-1)}$, which satisfies the condition
\begin{equation}\label{eq:periodic_equilibrium}
(\bI_n-\bA)z_0=\bB w_f .
\end{equation}
Conversely, every pair $(z_0,w_f)$
satisfying~\eqref{eq:periodic_equilibrium} defines an orbit with
states, inputs, and charges
\begin{equation}\label{eq:periodic_states}
z_i=A^iz_0+S_iQw_f,\quad u_i^f=Q_iw_f,\quad q_i^f=E_iQw_f,
\end{equation}
for $i=0,\ldots,h-1$. Let
$\xi_i^f:=[\,z_i^\top\ \ (q_i^f)^\top\,]^\top$.

Since charge balance is enforced over input blocks and $h$-CBC is formulated at block boundaries, we consider two cases. In Case I, the boundary state $z_0$ of the orbit is free. In Case II, it is fixed at $z_0=x_f$, i.e., $x_{ph}=x_f$. In both cases, our objective is to solve the optimal charge-balanced $h$-periodic orbits within the feasible set, and then design two types of controllers to regulate the system~\eqref{eq:sys} to the optimal charge-balanced $h$-periodic orbit.

Case I : we solve the optimal charge-balanced $h$-periodic orbit among all realizable charge-balanced $h$-periodic orbits that minimizes the deviation from $x_f$. To this end, we consider the single-cycle optimization problem
\begin{align}\label{eq:nonzero_case1}
\min_{z_0,w_f}\quad
J_{\rm I}
:=
\sum_{i=0}^{h-1}
\big(&
(z_i-x_f)^\top W_{x,i}(z_i-x_f)
+(u_i^f)^\top W_u u_i^f \notag\\
&+(q_i^f)^\top W_q q_i^f
\big)
\end{align}
subject to
\begin{equation}\label{eq:case1_constraint}
(\bI_n-\bA)z_0=\bB w_f.
\end{equation}
The optimal charge-balanced $h$-periodic orbit is
\begin{equation}\label{eq:case1_optimal_orbit}
z_i^\star=A^iz_0^\star+S_iQw_f^\star,~~~~~
u_i^{f\star}=Q_iw_f^\star,~~~~~
q_i^{f\star}=E_iQw_f^\star .
\end{equation}

Case II : a charge-balanced $h$-periodic orbit satisfying the $x_{ph}=x_f$ exists if and only if
\begin{equation}\label{eq:case2_condition}
(\bI_n-\bA)x_f\in\Span(\bB).
\end{equation}
Suppose~\eqref{eq:case2_condition} holds, consider the single-cycle optimization problem
\begin{align}\label{eq:nonzero_case2}
\min_{w_f}\quad
J_{\rm II}
:=
\sum_{i=0}^{h-1}
\big(&
(z_i-x_f)^\top W_{x,i}(z_i-x_f)
+(u_i^f)^\top W_u u_i^f \notag\\
&+(q_i^f)^\top W_q q_i^f
\big)
\end{align}
subject to
\begin{equation}\label{eq:case2_constraint}
(\bI_n-\bA)x_f=\bB w_f,
\end{equation}
where
\begin{equation*}
z_i=A^ix_f+S_iQw_f,\qquad
u_i^f=Q_iw_f,\qquad
q_i^f=E_iQw_f .
\end{equation*}

In the following, let
$\{z_i^\star,u_i^{f\star}\}_{i=0}^{h-1}$
denote the selected optimal charge-balanced $h$-periodic orbit obtained from either Case I or Case II. Next, we design two types of controllers to regulate the system~\eqref{eq:sys} to the optimal charge-balanced $h$-periodic orbit.

\subsection{Optimal periodic orbit step-wise regulation} 
For $k=ph+i$ with $i=0,\ldots,h-1$, define $\delta x_{k}:=x_k-z_i^\star$, $\delta q_{k}:=q_k-q_i^{f\star}$ and $\delta u_{k}:=u_k-u_i^{f\star}$.

Define the error state $\delta\xi_{k}:=[\,{\delta x_k}^\top\ \ {\delta q_k}^\top\,]^\top\in\R^{n+m}$, which evolves as
\begin{equation}\label{eq:error}
\delta\xi_{k+1}=\tilde A\delta\xi_k+\tilde B\delta u_k,
\end{equation}
where $\tilde A$ and $ \tilde B$ are the same matrices as in~\eqref{eq:aug}.

Hence, the system~\eqref{eq:error} has the same $h$-periodic augmented structure as the system~\eqref{eq:aug} considered in the origin stabilization problem. To this end, define the cost function as
\begin{align}\label{eq:Je}
J_{e}=\sum_{p=0}^{\infty}\sum_{i=0}^{h-1}
\big(&\delta x_{ph+i}^\top W_{x,i}\delta x_{ph+i}
+\delta u_{ph+i}^\top W_u\delta u_{ph+i}\notag\\
&+\delta q_{ph+i}^\top W_q\delta q_{ph+i}\big).
\end{align}
Since the deviation dynamics and the last-step charge-cancellation rule are identical in form to those of the origin-stabilization problem, the periodic Riccati equation and the gains $\tilde K_i$ derived previously apply directly. The corresponding step-wise regulation controller is
\begin{equation}\label{eq:reg_controller}
u_{ph+i}=
\begin{cases}
u_i^{f\star}-\tilde K_i\delta\xi_{ph+i}, & i=0,\ldots,h-2,\\
-q_{ph+h-1}, & i=h-1.
\end{cases}
\end{equation}
The controller consists of the optimal feedforward input $u_i^{f\star}$ and a step-wise feedback based on the deviation from the optimal charge-balanced $h$-periodic orbit. Thus, although system~\eqref{eq:sys} cannot be stabilized at a nonzero $x_f$ with charge-balanced inputs, the proposed controller~\eqref{eq:reg_controller} regulates the system~\eqref{eq:sys} to the optimal charge-balanced $h$-periodic orbit.

Since the system~\eqref{eq:error} has the same $h$-periodic augmented structure as the system~\eqref{eq:aug} considered in the origin stabilization problem, the stability result in Theorem~\ref{thm:step} applies directly. Thus, the optimal charge-balanced $h$-periodic orbit is globally asymptotically stable.
\subsection{Optimal periodic orbit cycle-wise regulation}
Let $U_f^\star=[\,u_0^{f\star\top}\ \cdots\ u_{h-1}^{f\star\top}\,]^\top$ with $RU_f^\star=\bZero$. For the $p$-th cycle, define the input-block error $\delta U_p:=U_p-U_f^\star$. Since both $U_p$ and $U^\star_f$ are charge-balanced, $R\delta U_p=\bZero$. Hence, there exists a $\delta w_p$ such that $\delta U_p=Q\delta w_p$. 

Consider a cycle-wise feedback controller
\begin{align}
    \delta w_p=K\delta x_{ph},\qquad \delta x_{ph}:=x_{ph}-z_0^\star.
\end{align}
Then, for $i=0,\ldots,h-1$, the optimal charge-balanced $h$-periodic orbit states satisfy
\begin{align}
z_{i}^\star=A^iz_{0}^\star+S_iU_f^\star.
\end{align}
Therefore,
\begin{align}\label{eq:sce}
\delta x_{ph+i}=A^i\delta x_{ph}+S_i\delta U_p
=(A^i+S_iQK)\delta x_{ph},
\end{align}
and the boundary states satisfy
\begin{align}\label{eq:fc}
\delta x_{(p+1)h}=F\delta x_{ph}
\end{align}
with the same lifted pair $(\bA,\bB)$ as in system~\eqref{eq:closed}. Thus, the system~\eqref{eq:fc} is  $h$-CBS. The optimal cycle-wise feedback gain $K^*$ is obtained from~\eqref{eq:optimal_K}. The cycle-wise regulation controller is
\begin{align}\label{eq:blo_controller}
U_p=U_f^\star+QK^*(x_{ph}-z_0^\star).
\end{align}
By Theorems~\ref{thm:step} and~\ref{thm:block}, $\lim_{p\to\infty}\delta x_{ph+i}=\bZero$. Hence, $\lim_{p\to\infty} x_{ph+i}=z_{i}^\star$.
\begin{thm}[Orbit tracking]\label{thm:tracking}
Suppose that system~\eqref{eq:sys} is $h$-CBS and that $(z_0,w_f)$
satisfies~\eqref{eq:periodic_equilibrium}. Then,
\begin{enumerate}
\item[\textup{(i)}] under step-wise regulation controller~\eqref{eq:reg_controller}, the origin of the
deviation dynamics~\eqref{eq:error} is globally asymptotically stable;
\item[\textup{(ii)}] under cycle-wise regulation controller~\eqref{eq:blo_controller}, the origin of
the deviation dynamics~\eqref{eq:sce} is globally asymptotically stable if and only
if $F=\bA+\bB K$ is Schur;
\item[\textup{(iii)}] with $K=K^*$, Theorems~\ref{thm:step} and \ref{thm:block} hold for the cost function~\eqref{eq:Je}.
\end{enumerate}
\end{thm}

\begin{figure}[t]
	\centering
	\includegraphics[width=0.9\columnwidth]{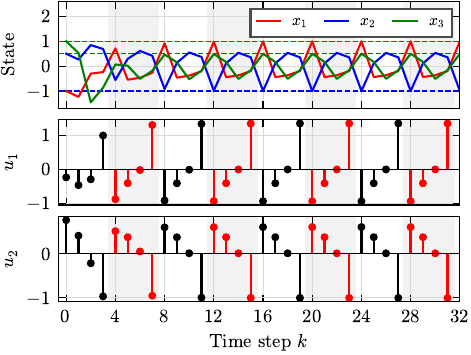}
	\caption{Stabilization of the selected orbit, with $x_f$ reached at
		the start of every cycle. Dashed lines mark the entries of $x_f$.}
	\label{fig:orbit}
\end{figure}

\begin{example}
    Consider system~\eqref{eq:sys} with $h=4$,
\begin{equation*}
A=\begin{bmatrix}-1&-2&-1\\ -1&-1&-1\\ 2&2&1\end{bmatrix},\qquad
B=\begin{bmatrix}1&0\\ 0&1\\ 1&1\end{bmatrix},
\end{equation*}
and the target $x_f=[1~\ -1~\ 0.5]^\top$, to be reached at the start
of every cycle, i.e., $x_{ph}=x_f$. The pair $(A,B)$ is
controllable with $\nu=2$, $1\notin\sigma(A)$, and $\sigma(A)$
contains no complete spectral $4$-cycle since $n<h$. Hence the system is
$4$-CBC by Theorem~\ref{thm:CBC}, and $x_{ph}=x_f$ is feasible since $h\ge\nu+1$. Problem~\eqref{eq:nonzero_case2} selects the Optimal charge-
balanced $h$-periodic orbit. The orbit is stabilized by the block
controller~\eqref{eq:blo_controller} with $K^*$ designed as in
Section~\ref{sec:origin} with $W_{x,i}=\bI_3$, $W_u=\bI_2$, and
$W_q=\bZero$. Fig.~\ref{fig:orbit} shows the trajectories from 
$x_0=[-1~\ 0.5~\ 1]^\top$ the boundary states $x_{4p}$ converge to
$x_f$, and the remaining states converge to the orbit.
\end{example}

\section{Application to Hodgkin-Huxley Control}\label{sec:hh}

We apply the cycle-wise controller to the Hodgkin--Huxley neuron model
\begin{align}\label{eq:hh}
  C_m\dot V&=I_{\rm app}+u-g_{\rm Na}m_{\rm Na}^3h_{\rm Na}(V-E_{\rm Na})
             \notag\\
           &\quad -g_{\rm K}n_{\rm K}^4(V-E_{\rm K})-g_{\rm L}(V-E_{\rm L}),\\
  \dot\zeta&=\alpha_\zeta(V)(1-\zeta)-\beta_\zeta(V)\zeta,\quad
          \zeta\in\{m_{\rm Na},h_{\rm Na},n_{\rm K}\},\notag
\end{align}
where $V$ is the membrane voltage in mV, $m_{\rm Na}$, $h_{\rm Na}$,
and $n_{\rm K}$ are gating variables with the standard rate functions
$\alpha_\zeta$ and $\beta_\zeta$, and
$x:=(V,m_{\rm Na},h_{\rm Na},n_{\rm K})^\top$. The parameters are
$C_m=1~\mu$F/cm$^2$, $g_{\rm Na}=120$, $g_{\rm K}=36$, and
$g_{\rm L}=0.3$~mS/cm$^2$, and $E_{\rm Na}=50$, $E_{\rm K}=-77$, and
$E_{\rm L}=-54.4$~mV. The constant current
$I_{\rm app}=20~\mu$A/cm$^2$ induces tonic spiking around the unique
equilibrium $x_{\rm e}$, with $V_{\rm e}\approx-56.59$~mV, which is
unstable, and $u$ is the stimulation current.

\begin{figure}[t]
	\centering
	\includegraphics[width=.9\columnwidth]{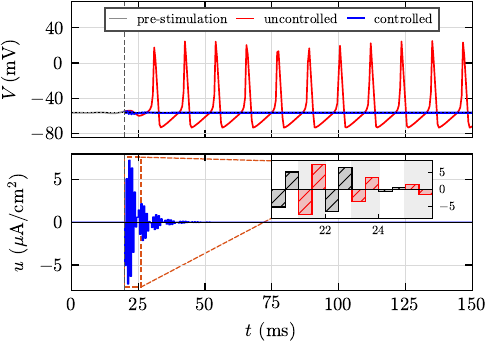}
	\caption{Stimulation starting at $t=20$~ms (dashed), before tonic
		spiking is established. Inset: the first cycles of the
		stimulation, with alternate cycles shaded.}
	\label{fig:hh_local}
\end{figure}

We linearize~\eqref{eq:hh} at $(x_{\rm e},0)$ and discretize it with a zero-order hold and sampling time $T_s=0.5$~ms. With $h=2$, the resulting  system~\eqref{eq:sys} is $2$-CBC by Theorem~\ref{thm:CBC}. Since $u$ is constant over each sampling interval,
$\int_{2pT_s}^{(2p+2)T_s}u(t)\,dt=T_s(u_{2p}+u_{2p+1})=0$, so the continuous-time stimulation is charge-balanced. The cycle-wise controller~\eqref{eq:lqr_original_controller} is designed in the scaled coordinates $D^{-1}(x-x_{\rm e})$ and $u/10$, with $D=\diag(10,0.1,0.1,0.1)$, $W_{x,0}=\bI_4$, $W_{x,1}=\bZero$, $W_u=0.3$, and $W_q=0$. It is applied to the nonlinear model~\eqref{eq:hh}, with the initial condition $x_0=x_{\rm e}+(0.1,0,0,0)^\top$.

\begin{figure}[t]
	\centering
	\includegraphics[width=.9\columnwidth]{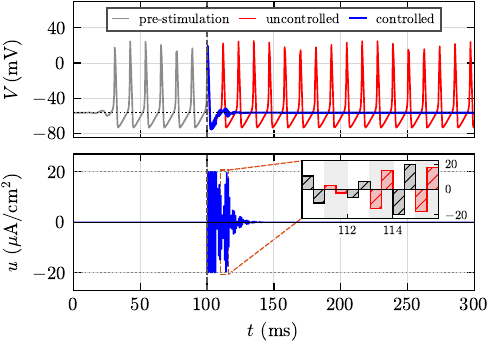}
	\caption{Stimulation starting at $t=100$~ms (dashed), after tonic
		spiking is established, with $|u|\le20~\mu$A/cm$^2$ (dotted).}
	% Inset as in Fig.~\ref{fig:hh_local}.}
\label{fig:hh_spiking}
\end{figure}

In Fig.~\ref{fig:hh_local}, stimulation starts at $t=20$~ms, before tonic spiking is established. The controlled state converges to $x_{\rm e}$, whereas the uncontrolled neuron spikes. In Fig.~\ref{fig:hh_spiking}, stimulation starts at $t=100$~ms, after tonic spiking is established, and is limited to $|u|\le20~\mu$A/cm$^2$ by saturating the block coordinate $w_p$, so that each pair remains balanced. The stimulation suppresses the tonic spiking and returns the state to $x_{\rm e}$, although it starts on the spiking regime, far from $x_{\rm e}$.

\section{Conclusion}\label{sec:con}
This article investigated the controllability, stabilization, and regulation of discrete-time linear systems subject to charge-balanced input constraints, motivated by electrical stimulation applications. We derived new necessary and sufficient conditions for charge-balanced controllability and stabilizability. These conditions characterize that the charge-balanced inputs reduce the effective input directions and prevent full-state steering when uncontrollable complete spectral $h$-cycles arise. In addition, we show that nonzero steady states cannot be held under charge-balanced inputs. Therefore, we designed optimal step-wise and cycle-wise feedback controllers that stabilize the origin as well as track optimal charge-balanced $h$-periodic orbits. We have applied our results to a nonlinear Hodgkin-Huxley neuron model, successfully suppressing the tonic spiking and returning the state to equilibrium. In the future, we are interested in extending the present discrete-time control framework under charge-balanced inputs to both continuous-time linear and nonlinear systems. 

\appendix
\section{Proofs in Section~\ref{sec:controllability}}\label{proof_sec_3}
\begin{pfof}{Lemma~\ref{lem:equiv} }
For a fixed $b$ the states reachable from $x_0$ at time $bh$ under \eqref{eq:cb}
are $\bA^bx_0+\CV_b$, where
$\CV_b:=\Span\big[\bA^{b-1}\bB\ \cdots\ \bA\bB\ \ \bB\big]
=\sum_{i=0}^{b-1}\bA^{i}\Span(\bB)$, because $U_p$ ranges over
$\Ker(R)=\Span(Q)$ exactly as $w_p$ ranges over $\R^{m(h-1)}$.  The subspaces
$\CV_b$ are nondecreasing in $b$ and stationary for $b\ge n$ by the
Cayley--Hamilton theorem.  Hence if $\CV_b=\R^n$ for some $b$, that single $b$
serves all pairs $(x_0,x_f)$; and if $\CV_b\neq\R^n$ for every $b$, then
$\CV_b\subseteq\CV_n\subsetneq\R^n$, so taking $x_0=\bZero$ and
$x_f\notin\CV_n$ shows the system is not $h$-CBC.  Thus $h$-CBC is equivalent to
$\CV_n=\R^n$, i.e.\ to controllability of $(\bA,\bB)$.  If $\tilde Q$ is another
basis matrix, then $\tilde Q=QT$ with $T$ invertible, so
$\Span(S\tilde Q)=\Span(SQ)$ and the reachable subspaces are identical.
\end{pfof}

\begin{pfof}{Lemma~\ref{lem:image} }
\emph{First equality.}  Since $\Span(Q)=\Ker(R)$ we have
$\Span(\bB)=S\,\Ker(R)$.  Partition $v\in\R^{mh}$ as
$v=(v_1^\top,\dots,v_h^\top)^\top$ with $v_r\in\R^m$; then $Rv=\sum_{r}v_r$ and
$Sv=\sum_{r=1}^{h}A^{h-r}Bv_r$.  The map
$\Phi:\R^{m(h-1)}\to\Ker(R)$ defined by $v_{h-j}:=w_j$ for $j=1,\dots,h-1$ and
$v_h:=-\sum_{j=1}^{h-1}w_j$ is a linear bijection onto $\Ker(R)$: it is well
defined since $Rv=\bZero$, clearly injective, and surjective because any
$v\in\Ker(R)$ arises from $w_j:=v_{h-j}$.  For $v=\Phi(w)$,
\[
  Sv=\sum_{r=1}^{h-1}A^{h-r}Bv_r+Bv_h
    =\sum_{j=1}^{h-1}(A^{j}-\bI_n)Bw_j
\]
after substituting $j=h-r$.  As $w$ ranges over $\R^{m(h-1)}$ the right-hand side
ranges over $\sum_{j=1}^{h-1}\Span\big((A^{j}-\bI_n)B\big)$.

\emph{Second equality.}  Since $A^{j}-\bI_n=(A-\bI_n)M_j$ with
$M_j:=\sum_{l=0}^{j-1}A^{l}$, it suffices to prove
$\mathcal W:=\sum_{j=1}^{h-1}\Span(M_jB)=\CK_{h-1}(A,B)$.  The inclusion $\subseteq$
holds because $M_jB$ is a combination of $B,AB,\dots,A^{j-1}B$ with $j-1\le h-2$.
Conversely $M_1B=B$, and $\mathcal W$ being a subspace containing $\Span(M_jB)$ and
$\Span(M_{j+1}B)$ also contains the image of $M_{j+1}B-M_jB=A^{j}B$ for
$j=1,\dots,h-2$; hence $\CK_{h-1}(A,B)\subseteq\mathcal W$.
\end{pfof}

\begin{pfof}{Lemma~\ref{lem:necessary} }
By Lemma~\ref{lem:equiv}, $h$-CBC implies that $(\bar A,\bar B)$ is controllable. Suppose that $(A,B)$ is uncontrollable. By the PBH test, there exist $\lambda\in\BC$ and $\phi\neq\bZero$ such that $\phi^\top A=\lambda\phi^\top$ and $\phi^\top B=\bZero^\top$. Hence $\phi^\top A^h=\lambda^h\phi^\top$ and $\phi^\top A^jB=\lambda^j\phi^\top B=\bZero^\top$ for all $j\ge0$, which gives $\phi^\top\bar B=\bZero^\top$. This violates the PBH test for $(\bar A,\bar B)$, and therefore $(A,B)$ is controllable. Next, suppose that $1\in\sigma(A)$. Then there exists $\phi\neq\bZero$ such that $\phi^\top A=\phi^\top$. It follows that $\phi^\top A^h=\phi^\top$ and $\phi^\top(A^j-\bI_n)B=\bZero^\top$ for all $j=1,\ldots,h-1$. By Lemma~\ref{lem:image}, $\phi^\top\bar B=\bZero^\top$, again contradicting the PBH test for $(\bar A,\bar B)$. Thus, $1\notin\sigma(A)$.
\end{pfof}

\begin{pfof}{Lemma~\ref{lem:lift} }
Write $M:=A^\top$ and decompose
$\BC^n=\oplus_{\lambda\in\sigma(M)}\CG_\lambda$ into generalized eigenspaces, noting $\sigma(M)=\sigma(A)$.  Each $\CG_\lambda$ is $M$-invariant, hence $(M^h-\mu \bI_n)$-invariant, so writing $\phi=\sum_\lambda\phi_\lambda$ we have
$(M^h-\mu \bI_n)\phi=\bZero$ if and only if $(M^h-\mu \bI_n)\phi_\lambda=\bZero$ for every $\lambda$; it suffices to treat
each generalized eigenspace.  Fix $\lambda$ and write $M|_{\CG_\lambda}=\lambda \bI+N$ with $N$ nilpotent.

If $\lambda=0$, then $M|_{\CG_0}=N$ is nilpotent, so $M^h|_{\CG_0}$ is nilpotent
and, as $\mu\neq0$, $M^h-\mu \bI$ is invertible on $\CG_0$; hence
$\phi_0=\bZero$.

If $\lambda\neq0$, the binomial theorem gives
$M^h|_{\CG_\lambda}=\lambda^h\bI+N'$ with
$N':=\sum_{k=1}^{h}\binom{h}{k}\lambda^{h-k}N^{k}$ nilpotent, so
$M^h-\mu \bI=(\lambda^h-\mu)\bI+N'$ on $\CG_\lambda$.  If $\lambda^h\neq\mu$
this is a nonzero scalar plus a nilpotent, hence invertible, and
$\phi_\lambda=\bZero$.  If $\lambda^h=\mu$ the condition becomes
$N'\phi_\lambda=\bZero$.  Factor $N'=P(N)N$ with
$P(N):=h\lambda^{h-1}\bI+\sum_{k=2}^{h}\binom{h}{k}\lambda^{h-k}N^{k-1}$, which
is legitimate because $N$ and $P(N)$ commute.  Since $\lambda\neq0$, $P(N)$ is a
nonzero multiple of the identity plus a nilpotent, hence invertible, so
$N'\phi_\lambda=\bZero\iff N\phi_\lambda=\bZero\iff
\phi_\lambda\in\Ker(M-\lambda \bI)$.  Collecting the three cases
yields~\eqref{eq:lifteig}; the sum is direct because the $\CG_\lambda$ are
independent.
\end{pfof}

\begin{pfof}{Lemma~\ref{lem:zero}}
Suppose such a $\phi$ exists.  If $A$ is nonsingular then $\nu_0=0$ and
$\phi^\top A^h=\bZero^\top$ already forces $\phi=\bZero$; so assume
$\nu_0\ge1$.  By Lemma~\ref{lem:image}, $\phi^\top\bB=\bZero^\top$ is equivalent
to
\begin{equation}\label{eq:zerocond}
  \phi^\top A^{j}B=\phi^\top B,\qquad j=1,\dots,h-1 .
\end{equation}
Since $h\ge\nu_0+1$, stabilization of the kernel chain gives
$\Ker\big((A^\top)^h\big)=\Ker\big((A^\top)^{\nu_0}\big)$. Thus $\phi^\top A^h=\bZero^\top$ implies
$\phi^\top A^{\nu_0}=\bZero^\top$. Taking $j=\nu_0$ in~\eqref{eq:zerocond}, which is admissible since $1\le\nu_0\le h-1$, gives
$\phi^\top B=\phi^\top A^{\nu_0}B=\bZero^\top$; substituting back yields $\phi^\top A^{j}B=\bZero^\top$ for $j=0,\dots,h-1$, while
$\phi^\top A^{j}=\bZero^\top$ for $j\ge\nu_0$.
Hence $\phi^\top A^jB=\bZero^\top$ for all $j\ge0$, so $\phi^\top\CK_n(A,B)=\{0\}$ with $\phi\neq\bZero$, contradicting controllability of $(A,B)$.
\end{pfof}

\begin{pfof}{Lemma~\ref{lem:partialcycle}}
Let $\Lambda_\mu=\{\lambda_1,\ldots,\lambda_s\}$ with $s\le h-1$, and suppose that there exists $\phi\neq\bZero$ such that $\phi^\top A^h=\mu\phi^\top$ and $\phi^\top\bar B=\bZero^\top$. By Lemma~\ref{lem:lift}, $\phi$ admits the unique decomposition $\phi=\sum_{i=1}^{s}\phi_i$, where $\phi_i\in\Ker(A^\top-\lambda_i \bI_n)$. Set $c_i:=\phi_i^\top B$ and $c:=\sum_{i=1}^{s}c_i$. By Lemma~\ref{lem:image}, $\phi^\top\bar B=\bZero^\top$ is equivalent to
\[
    \sum_{i=1}^{s}(\lambda_i^j-1)c_i=\bZero^\top,
    \qquad j=1,\ldots,h-1,
\]
which also holds for $j=0$. Let $v(z):=(1,z,\ldots,z^{h-1})^\top\in\BC^h$, $V:=[\,v(\lambda_1)\ \cdots\ v(\lambda_s)\,]$, and let $C\in\BC^{s\times m}$ have $c_i$ as its $i$th row. Then $VC=v(1)c$, $[\,V\ \ v(1)\,]\,[\,C^\top\ -c^\top]^\top=\bZero$. Since $1\notin\sigma(A)$, the nodes $\lambda_1,\ldots,\lambda_s,1$ are distinct, and $s+1\le h$. Hence $[\,V\ \ v(1)\,]$ is a Vandermonde matrix with full column rank, which gives $C=\bZero$. Thus $\phi_i^\top B=\bZero$ for all $i$. Since $(A,B)$ is controllable, the PBH test implies $\phi_i=\bZero$ for all $i$, and hence $\phi=\bZero$, a contradiction.
\end{pfof}

\begin{pfof}{Lemma~\ref{lem:completecycle}}
Let $\Lambda_\mu=\{\lambda_1,\ldots,\lambda_h\}$.

\emph{Necessity.}
Suppose that there exists $\phi\neq\bZero$ satisfying
$\phi^\top A^h=\mu\phi^\top$ and
$\phi^\top\bar B=\bZero^\top$. By Lemma~\ref{lem:lift},
$\phi=\sum_{i=1}^{h}\phi_i$ uniquely, where
$\phi_i\in\Ker(A^\top-\lambda_i \bI_n)$. Set
$c_i:=\phi_i^\top B$ and $c:=\sum_{i=1}^{h}c_i$. By
Lemma~\ref{lem:image},
$\sum_{i=1}^{h}(\lambda_i^j-1)c_i=\bZero$ for
$j=0,\ldots,h-1$. Let
$v(z):=(1,z,\ldots,z^{h-1})^\top$,
$V:=[\,v(\lambda_1)\ \cdots\ v(\lambda_h)\,]$, and let $C$ have
$c_i$ as its $i$th row. Since the $\lambda_i$ are distinct, $V$ is
invertible, and $VC=v(1)c$ gives $c_i=\gamma_i c$, where
$\gamma:=V^{-1}v(1)$.
Since the $\lambda_i$ are the distinct $h$th roots of $\mu$,
the ratio $\lambda_k/\lambda_i$ is an $h$th root of unity, equal to
$1$ exactly when $k=i$. Thus, for $i,k=1,\ldots,h$,
\[
\frac1h\sum_{l=0}^{h-1}\left(\frac{\lambda_k}{\lambda_i}\right)^l
=\begin{cases}1,&k=i,\\0,&k\neq i.\end{cases}
\]
It follows that $(V^{-1})_{i,l+1}=h^{-1}\lambda_i^{-l}$ for
$l=0,\ldots,h-1$. Multiplying $V^{-1}$ by $v(1)$ gives
\begin{equation}\label{eq:gamma}
\gamma_i=\frac1h\sum_{l=0}^{h-1}\lambda_i^{-l}
          =\frac1h\cdot\frac{1-\mu^{-1}}{1-\lambda_i^{-1}}\ \neq\ 0 .
\end{equation}
Since $1\notin\sigma(A)$, $\lambda_i\neq1$ for every $i$. Also,
$\mu\neq1$, because otherwise the complete set of $h$th roots of
$\mu$ would contain $1$. Hence $\gamma_i\neq0$ for every $i$.

If $c=\bZero$, then $c_i=\bZero$ for all $i$, and controllability of $(A,B)$
together with the PBH test gives $\phi_i=\bZero$ for all $i$, contradicting
$\phi\neq\bZero$. Thus $c\neq\bZero$. Since
$c_i=\gamma_i c\in\CB_{\lambda_i}$ and $\gamma_i\neq0$, it follows that
$c\in\CB_{\lambda_i}$ for every $i$. Therefore,
$\bigcap_{\lambda\in\Lambda_\mu}\CB_\lambda\neq\{\bZero\}$.

\emph{Sufficiency.}
Suppose that
$\bigcap_{\lambda\in\Lambda_\mu}\CB_\lambda\neq\{\bZero\}$ and choose
$c\neq\bZero$ in this intersection. Let $\gamma_i$ be defined by
\eqref{eq:gamma}. Since $\gamma_i c\in\CB_{\lambda_i}$, there exists
$\phi_i\in\Ker(A^\top-\lambda_i \bI_n)$ such that
$\phi_i^\top B=\gamma_i c$. Because $\gamma_i c\neq\bZero$, each $\phi_i$
is nonzero. Since these vectors belong to eigenspaces associated with
distinct eigenvalues, $\phi:=\sum_{i=1}^{h}\phi_i\neq\bZero$. Moreover,
$\lambda_i^h=\mu$ implies $\phi^\top A^h=\mu\phi^\top$.
The identity $V\gamma=v(1)$ gives
$\sum_{i=1}^h\gamma_i\lambda_i^j=1$ for $j=0,\ldots,h-1$;
in particular, $\sum_{i=1}^h\gamma_i=1$.
Finally, for
$j=0,\ldots,h-1$,
\[
  \phi^\top(A^{j}-\bI_n)B
  =\Big(\sum_i\gamma_i\lambda_i^{\,j}-\sum_i\gamma_i\Big)c=\bZero^\top.
\]
By Lemma~\ref{lem:image}, $\phi^\top\bar B=\bZero^\top$. Hence there
exists a nonzero $\phi\in\Ker((A^\top)^h-\mu \bI_n)$ satisfying
$\phi^\top\bar B=\bZero^\top$.
\end{pfof}

\section{Proofs in Section~\ref{sec:cbs}}\label{proof_sec_4}

\begin{pfof}{Lemma~\ref{lem:cbs_equiv}}
Every charge-balanced input block has the form $U_p=Qw_p$,
so the boundary states satisfy~\eqref{eq:lift}.
If $(\bA,\bB)$ is not stabilizable, the PBH test gives
$\phi\neq\bZero$ and $|\mu|\ge1$ such that, under every
admissible input sequence,
\[
\phi^\top x_{ph}=\mu^p\phi^\top x_0.
\]
Choosing a real $x_0$ with $\phi^\top x_0\neq0$ rules out
$x_{ph}\to\bZero$, contradicting $h$-CBS.
Conversely, if $(\bA,\bB)$ is stabilizable, choose $K$ such
that $F:=\bA+\bB K$ is Schur and apply $U_p=QKx_{ph}$.
Then $x_{ph}=F^px_0\to\bZero$. By~\eqref{d10}, each
intra-cycle state is a fixed linear function of $x_{ph}$.
Since there are finitely many phases, $x_k\to\bZero$.
Basis independence follows from the same invertible input
transformation used in the proof of Lemma~\ref{lem:equiv},
which also preserves stabilizability.
\end{pfof}

\begin{pfof}{Theorem~\ref{thm:CBS}}
\emph{Necessity.}
Suppose that system~\eqref{eq:sys} is $h$-CBS. By
Lemma~\ref{lem:cbs_equiv}, $(\bA,\bB)$ is stabilizable.
If \textup{(i)} fails, the PBH test gives $\lambda$ with
$|\lambda|\ge1$ and $\phi\neq\bZero$ such that
$\phi^\top A=\lambda\phi^\top$ and $\phi^\top B=\bZero^\top$.
Then $\phi^\top A^h=\lambda^h\phi^\top$ and
$\phi^\top A^jB=\bZero^\top$ for all $j\ge0$, so
$\phi^\top\bB=\phi^\top SQ=\bZero^\top$.
Since $|\lambda^h|\ge1$, this contradicts stabilizability of the
lifted pair. Thus \textup{(i)} holds.
If \textup{(ii)} fails, there is $\phi\neq\bZero$ with
$\phi^\top A=\phi^\top$. Lemma~\ref{lem:image} gives
$\phi^\top\bB=\bZero^\top$, while $\phi^\top A^h=\phi^\top$,
again contradicting stabilizability of the lifted pair.
Thus \textup{(ii)} holds.
If \textup{(iii)} fails, there is a complete spectral $h$-cycle
$\Lam_\mu\subseteq\sigma(A)$ with $|\mu|\ge1$ and
$\bigcap_{\lambda\in\Lam_\mu}\CB_\lambda\neq\{\bZero\}$.
The sufficiency argument in the proof of
Lemma~\ref{lem:completecycle}, which does not use controllability
of $(A,B)$, then yields $\phi\neq\bZero$ with
$\phi^\top A^h=\mu\phi^\top$ and $\phi^\top\bB=\bZero^\top$.
This contradicts stabilizability of the lifted pair, so
\textup{(iii)} holds.

\emph{Sufficiency.}
Assume \textup{(i)}--\textup{(iii)}. By Lemma~\ref{lem:cbs_equiv},
it suffices to show that $(\bA,\bB)$ is stabilizable.
Suppose otherwise. The PBH test gives $\mu$ with $|\mu|\ge1$ and
$\phi\neq\bZero$ such that $\phi^\top A^h=\mu\phi^\top$ and
$\phi^\top\bB=\bZero^\top$.
Let $\Lam_\mu:=\{\lambda\in\sigma(A):\lambda^h=\mu\}$ and
$s:=|\Lam_\mu|$. Since $\mu\in\sigma(A^h)$, $1\le s\le h$.
Every $\lambda\in\Lam_\mu$ satisfies
$|\lambda|=|\mu|^{1/h}\ge1$, so \textup{(i)} gives
\[
\Rank[\,A-\lambda\bI_n\ \ B\,]=n,
\qquad \lambda\in\Lam_\mu.
\]
For a fixed $\mu\neq0$, the proofs of
Lemmas~\ref{lem:partialcycle} and~\ref{lem:completecycle} use
controllability only through these rank conditions on $\Lam_\mu$.
Together with \textup{(ii)}, they therefore allow the same arguments
to be applied here. If $s\le h-1$, the argument of
Lemma~\ref{lem:partialcycle} gives a contradiction.
If $s=h$, then $\Lam_\mu$ is a complete spectral $h$-cycle, and the
argument of Lemma~\ref{lem:completecycle} gives
$\bigcap_{\lambda\in\Lam_\mu}\CB_\lambda\neq\{\bZero\}$,
contradicting \textup{(iii)}.
Thus $(\bA,\bB)$ is stabilizable, and Lemma~\ref{lem:cbs_equiv}
implies that system~\eqref{eq:sys} is $h$-CBS.
\end{pfof}

\section{Proofs in  Sections~\ref{sec:origin} and~\ref{sec:nonzero}}\label{proof_sec_5_6}
\begin{pfof}{Theorem~\ref{thm:step}}
The proof is organized four steps: we first show that the one-cycle backward sweep realizes the lifted Riccati operator (Step~1), and then apply standard LQR theory to the lifted system to establish the uniqueness of the periodic solution (Step~2), optimality of the step-wise controller (Step~3), and global asymptotic stability (Step~4).

\emph{Step 1: one-cycle problem.} Since $W_u\succ0$ and $Q$ has
orthonormal columns, ${ \bar W_w}\succeq Q^\top(\bI_h\otimes W_u)Q\succ0$.
The function $\ell(x,w)$ in~\eqref{eq:lifted_cost} is the cost of one
cycle, so it is nonnegative, and its phase-$0$ term gives
$\ell(x,w)\ge x^\top W_{x,0}x$ for all $w$. Hence
${ \bar W_x}-{ \bar W_{xw}}{ \bar W_w}^{-1}{ \bar W_{xw}}^\top$, whose quadratic form equals
$\min_w\ell(x,w)$, satisfies
${ \bar W_x}-{ \bar W_{xw}}{ \bar W_w}^{-1}{ \bar W_{xw}}^\top\succeq W_{x,0}\succ0$. For $P\succeq0$, let
$\mathcal R(P)$ denote the right-hand side of~\eqref{eq:lqr_dare} and
$K_P:=-({ \bar W_w}+\bB^\top P\bB)^{-1}(\bB^\top P\bA+{ \bar W_{xw}}^\top)$. Completing
the square gives, for all $x$ and $w$,
\begin{align}\label{eq:cs_lifted}
&\ell(x,w)+(\bA x+\bB w)^\top P(\bA x+\bB w)\notag\\
&=x^\top\mathcal R(P)x+\|w-K_Px\|_{{ \bar W_w}+\bB^\top P\bB}^2 .
\end{align}
Consider the problem of minimizing the cost of one cycle plus the
terminal cost $x_{(p+1)h}^\top Px_{(p+1)h}$ over all
$U_p\in\Ker(R)$, for a given $x:=x_{ph}$. By~\eqref{eq:cs_lifted},
its minimum is $x^\top\mathcal R(P)x$ and its unique minimizer is
$w_p=K_Px$. In terms of $\xi$, the same problem has the dynamics
\eqref{eq:aug} at phases $0,\ldots,h-2$ and $\tilde A_{\rm e}$ at
phase $h-1$, the stage costs of Section~\ref{sec:origin}, the terminal
cost $\xi^\top\diag(P,\bZero)\xi$, and the initial state
$\xi_{ph}=[\,x^\top\ \ \bZero\,]^\top$.
Bellman's principle gives quadratic value functions
$\xi^\top\Pi_i\xi$ satisfying~\eqref{eq:sweep}, with
$\diag(P,\bZero)$ in place of $\Pi_0$ in its first line.
The nonnegative costs imply $\Pi_i\succeq0$, hence $\Theta_i\succ0$.
The minimizing feedback has the form~\eqref{eq:step_controller}
with these matrices, and its value at the cycle boundary is
$x^\top[\Pi_0]_xx$. Comparing with~\eqref{eq:cs_lifted} gives
\begin{equation}\label{eq:sweep_lift}
[\Pi_0]_x=\mathcal R(P).
\end{equation}
By uniqueness of the one-cycle minimizer, this feedback applies
$u_{ph+i}=Q_iK_Px$, $i=0,\ldots,h-1$.

\emph{Step 2: proof of \textup{(i)}.}
Since the last-step transition has zero charge,
\eqref{eq:sweep} determines all the phase matrices uniquely from
$P:=[\Pi_0]_x$. By~\eqref{eq:sweep_lift}, consistency requires
$P=\mathcal R(P)$. Thus, Step~1 gives a one-to-one correspondence
between the positive semidefinite solutions of~\eqref{eq:sweep}
and~\eqref{eq:lqr_dare}. With
$v_p:=w_p+{ \bar W_w}^{-1}{ \bar W_{xw}}^\top x_{ph}$, \eqref{eq:lqr_dare} is the
Riccati equation of a standard LQR problem for the pair
$(\bA-\bB { \bar W_w}^{-1}{ \bar W_{xw}}^\top,\bB)$, which is stabilizable because
$(\bA,\bB)$ is (Lemma~\ref{lem:cbs_equiv}), with positive definite state
weight ${ \bar W_x}-{ \bar W_{xw}}{ \bar W_w}^{-1}{ \bar W_{xw}}^\top$ and input weight ${ \bar W_w}$.
Hence~\eqref{eq:lqr_dare} has a unique positive semidefinite
solution $P$, which is stabilizing.
The correspondence above therefore yields a unique positive
semidefinite solution $\{\Pi_i\}_{i=0}^{h-1}$ of~\eqref{eq:sweep},
proving \textup{(i)}.

\emph{Step 3: proof of \textup{(ii)}.}
For the stabilizing solution $P$ in Step~2, standard LQR theory
gives the minimum $x_0^\top Px_0$ of~\eqref{eq:lifted_cost}, attained
by $w_p=K_Px_{ph}$. By Step~1, the step-wise controller generates
$U_p=QK_Px_{ph}$. Since $U_p=Qw_p$ parametrizes all charge-balanced
input blocks and~\eqref{eq:lifted_cost} equals the original cost,
this controller achieves the minimum $x_0^\top[\Pi_0]_xx_0$,
proving \textup{(ii)}.

\emph{Step 4: proof of \textup{(iii)}.}
By Step~1, the boundary dynamics are
$x_{(p+1)h}=Fx_{ph}$, where $F:=\bA+\bB K_P$ is Schur by Step~2.
Equation~\eqref{d10} gives
$x_{ph+i}=(A^i+S_iQK_P)F^px_0$.
Since there are finitely many phases, there exist $c>0$ and
$\rho\in(0,1)$ such that
$\|x_{ph+i}\|\le c\rho^p\|x_0\|$ for all $p$ and $i$.
This proves global asymptotic stability. 
\end{pfof}

\begin{pfof}{Theorem~\ref{thm:block}}
Step~2 of the proof of Theorem~\ref{thm:step} establishes the
unique stabilizing solution $P\succeq0$ of~\eqref{eq:lqr_dare}.
Standard LQR theory gives the optimal lifted feedback
$w_p=K^*x_{ph}$ with value $x_0^\top Px_0$, proving \textup{(i)}.
Since $U_p=Qw_p$ parametrizes all charge-balanced input blocks
and~\eqref{eq:lifted_cost} equals the original cost, the cycle-wise
controller achieves the minimum of~\eqref{eq:lqr_cost},
proving \textup{(ii)}.
Finally, $F:=\bA+\bB K^*$ is Schur, and~\eqref{d10} gives
$x_{ph+i}=(A^i+S_iQK^*)F^px_0$. The finitely many phase matrices
therefore imply global asymptotic stability.
\end{pfof}

\begin{pfof}{Theorem~\ref{thm:nohold}}
    Summing $x_{k+1}=Ax_k+Bu_k$ over $k=ph,\ldots,(p+1)h-1$ and
using~\eqref{eq:cb} gives
$\sum_{i=1}^{h}x_{ph+i}=A\sum_{i=0}^{h-1}x_{ph+i}$, which
is~\eqref{eq:cycle_sum}. If $x_k\to x_f$, letting $p\to\infty$
in~\eqref{eq:cycle_sum} gives $h(\bI_n-A)x_f=\bZero$, so
$x_f=\bZero$ because $1\notin\sigma(A)$.
\end{pfof}

\begin{pfof}{Theorem~\ref{thm:tracking}}
Since the deviation systems under step-wise and cycle-wise control are identical to their respective origin-stabilization counterparts~\eqref{eq:aug}-\eqref{eq:step_controller} and~\eqref{eq:closed}-\eqref{eq:lifted_cost}, statements \textup{(i)}-\textup{(iii)} follow directly from Theorems~\ref{thm:step} and~\ref{thm:block}. 
\end{pfof}

%===============================================================================
%\begin{ack}
%The authors thank \dots
%\end{ack}

%\bibliographystyle{plain}
\bibliographystyle{autosam.bst}
\bibliography{autosam}

\end{document}